\documentstyle[12pt]{article}
\input epsf
\begin{document}

{\hfill RU-97-38}

\begin{center}
  \Large Renormalizing Recitation Grades\\
  Joel Shapiro\\
  \large Department of Physics and Astronomy\\
  Rutgers University, Piscataway, NJ 08855-0849\\
  May 20, 1997
\end{center}

\bigskip
\begin{center}
  \large Abstract
\end{center}

I discuss issue of how to adjust recitation grades given by different
instructors in a large course, taking into account and correcting for
differences in standards among the instructors, while preserving the
effects of differences in average student performance among the
recitation sections. 

\bigskip
\bigskip
Introductory physics in large universities is dominantly
taught in large courses which combine lectures with recitation
sections. The course content and policy is set by one person who is 
the course leader, but there are a number of people responsible for
the recitation sections. The students receive grades based primarily
but not entirely on examinations. These are graded in a manner so that
variations in standards among the recitation instructors do not
unfairly affect some students in comparison to others.\footnote{Many of our
  examinations in these courses are computer-graded multiple choice,
  but even when the questions are graded in a more subjective fashion,
  a high level of evenhandedness can be assured by having each 
  instructor grade all the exams for a given question.}

That component of the grade which is based on recitation work,
however, cannot easily be assigned in a way that is fair to all students. 
Quizzes are of necessity different, and graded by people with widely differing
standards.  
Even if
the recitation grade is only a small part of the total,
so that the effect of these differences on a student's total
grade is not very significant, nonetheless the effect on the morale
of students is quite significant. They are acutely aware of
discrepancies in standards.
This paper addresses the issue of how to deal with this problem so that
both the students and instructors are confident that grades are being
fairly assigned. I call this the problem of renormalizing
recitation grades.

Before embarking on a convoluted mechanism for renormalizing the
grades, we need to be convinced that the perception of unfairness
is the result of a {\em real} problem. 
Instructors have very differing grading standards, even
after they have been asked to grade to a certain
standard. For example, in the course I have recently administered,
Prof.~A had two sections with an average exam grade of 49, 
while Prof.~B had four sections with an average exam grade of 48. 
Nonetheless the average recitation for Prof.~A was 65, 
while that of Prof.~B was 79. The students were very aware of this 
non-negligible difference and were quite upset, until
I told them that we knew how to correct for these differences, and those
with the tougher instructor would not be unfairly graded in the end. 
Unfortunately, I was stretching the truth --- at the time,
we did not really have such a
scheme in place. This is my attempt at developing one.

The problem is also {\em non-trivial}.
We could just add 14 points to each of Prof.~A's students, 
but no doubt this would push some of his students over 100, 
which would be noticeable and unfair to the best students in
Prof.~B's classes, as they would have no chance to get over 100. 
This simple approach also
ignores any inherent differences in the average ability of students among
different sections.
Most of us who have been involved in such courses
realize there are significant variations in the average ability among
the sections. For example,
Prof.~C in my
course had one section with an exam average of 44\% and another 
with an average of 51\%, a very significant difference, especially
considering that these were multiple choice exams where random
guessing alone provided 20\%.

  Since the beginning of the use of recitation sections and
computerized exams in our Department nearly twenty years ago, 
the necessity of dealing with
this problem has been recognized. But I believe the issues have not 
been well understood, much less solved, before now.

First, I would like to make clear that there are two essentially
independent issues here, once we agree to do some renormalization. 
Having decided to set a goal for the average renormalized recitation
grades for each section, the first issue is how
to {\em determine} that goal\footnote{Some might be more demanding and
  require not only a goal average but some other constraint on the
  distribution, such as a given standard deviation. I am
  not convinced that this is an important issue, however, and I will
  ignore it.}.
The other problem is 
what function $f$ to apply to the raw recitation grades $x_i$ to get
the new recitation grades $f(x_i)$ which will have an average which
{\em achieves} the goal. 
Neither has been done well in our Department in the past.

\medskip
\noindent{\bf Determining the Goal}

As stated above, recitation sections are far from
random samplings from the total class ensemble. The real reasons are
not known to me, but plenty of possible causes abound, in particular
conflict with other courses. For example, a recitation section that
meets at the same time as an honors calculus course might well have
students of lower average ability than the class as a whole.

In our Department, some course leaders have chosen to ignore such
differences in the sections and to normalize recitation grades by 
bringing each section to the same predetermined average.
Those who have tried to make individual
renormalization goals for each section have, to my knowledge, all used
the assumption that the recitation grade average should be
proportional to the exam averages. Thus
the goal $g_r$ is given by 
$ g_r = \left.\langle e_i \rangle_r G\right/\langle e_i \rangle_c,$
where the averages\footnote{I am using throughout 
  the notation $\langle v_i \rangle_S$ for the average of the data
  $v_i$ over some set $S$.}
of exam scores for the recitation section
$\langle e_i \rangle_r$ and for the whole class $\langle e_i \rangle_c$
are used to scale the overall class goal $G$.

\medskip
\noindent
\begin{minipage}[t]{3.0in}
\parindent=1.7em

The assumption that recitation grades are, in some average sense, 
{\bf proportional} to exam grades does not stand up
to inspection. In general exam distributions are roughly gaussian with
a fairly low mean, while recitation grades are distributed quite 
differently, with a large fraction of the distribution narrowly
clustered near a perfect grade, and a smaller wide tail.
In my course of over 400 students, a scatter plot of 
raw recitation grade versus exam grade shows a very wide dispersion,
but if one imposes a linear fit, roughly half of the average
recitation grade is due to the intercept.  Thus by assuming
proportionality one would be overcorrecting by roughly a factor of two.
One way to see how inappropriate this goal calculation
is is with the following fictional but still reasonable
\end{minipage}
\mbox{\ }
\begin{minipage}[t]{2.1in}
\leavevmode{\tiny\phantom{.}}\\[-5pt]
\epsfxsize=2.1in\epsfbox{xy.eps}\\
Scatter plot of recitation grade versus total exam grade in Spring,
1997. The best least squares linear fit is shown. Note that about half
of the recitation grade is given by the intercept and half by the
linear term.
\end{minipage}\\[3pt]
situation: 
Suppose a good section has an exam average of 70\% in a course where
the overall average is 55\%. Suppose the overall recitation grade goal is
80\%. This gives a goal for the average recitation grade in this
section of 102\%. Clearly wrong!

\medskip
\noindent {\bf The renormalization function}

The other issue, also not trivial, is this: Given a set
of raw recitation scores $\{x_i\}$ for a section and a goal $g$ 
that we
wish the average of the renormalized scores to be, 
how do we find a suitable function
$f$ for which $\langle f(x_i) \rangle = g$. Unfortunately, $f$ needs to
have some other suitable properties\footnote{For simplicity,
  throughout this discussion I will take
  the recitation grades to range from 0 to 100.}
\begin{description}
\item[1:] $f(0) = 0$.
\item[2:] $f(100) = 100$.
\end{description}
One should not be too quick to think he has a solution to this
problem. Clearly a simple scaling of the grades to adjust the average
to the goal, $f(x) = g x/\langle x_i\rangle$, does not meet requirement 2.
A piecewise linear fit, say with one segment from (0,0) to some 
intermediate point and another from that point to (100,100), can be
made to have the right average, but one must remember that the average
of the function is not the function applied to the average, and the
intermediate point is not determined by only $g$ and $\langle
x_i\rangle$. A quadratic fit is easily calculable;
$f(x)=(1-100 a)x + ax^2$ satisfies (1) and (2)
automatically. Getting the right average is simple:
\begin{minipage}[t]{3.0in}
\parindent=1.7em

$$a = {g-\langle x_i \rangle \over \langle x_i^2 \rangle 
- 100\langle x_i \rangle}.$$
However, it violates a rather serious requirement:
\begin{description}
\item[3:] $f$ must increase monotonically on the interval $[0,100]$,
\end{description}
in some quite ordinary situations, including the first recitation
section I taught in which this method was being employed.

\end{minipage}
\mbox{\ }
\begin{minipage}[t]{2.0in}
    \leavevmode\hbox{\phantom{a}}\\
\epsfxsize=2.0in\epsfbox{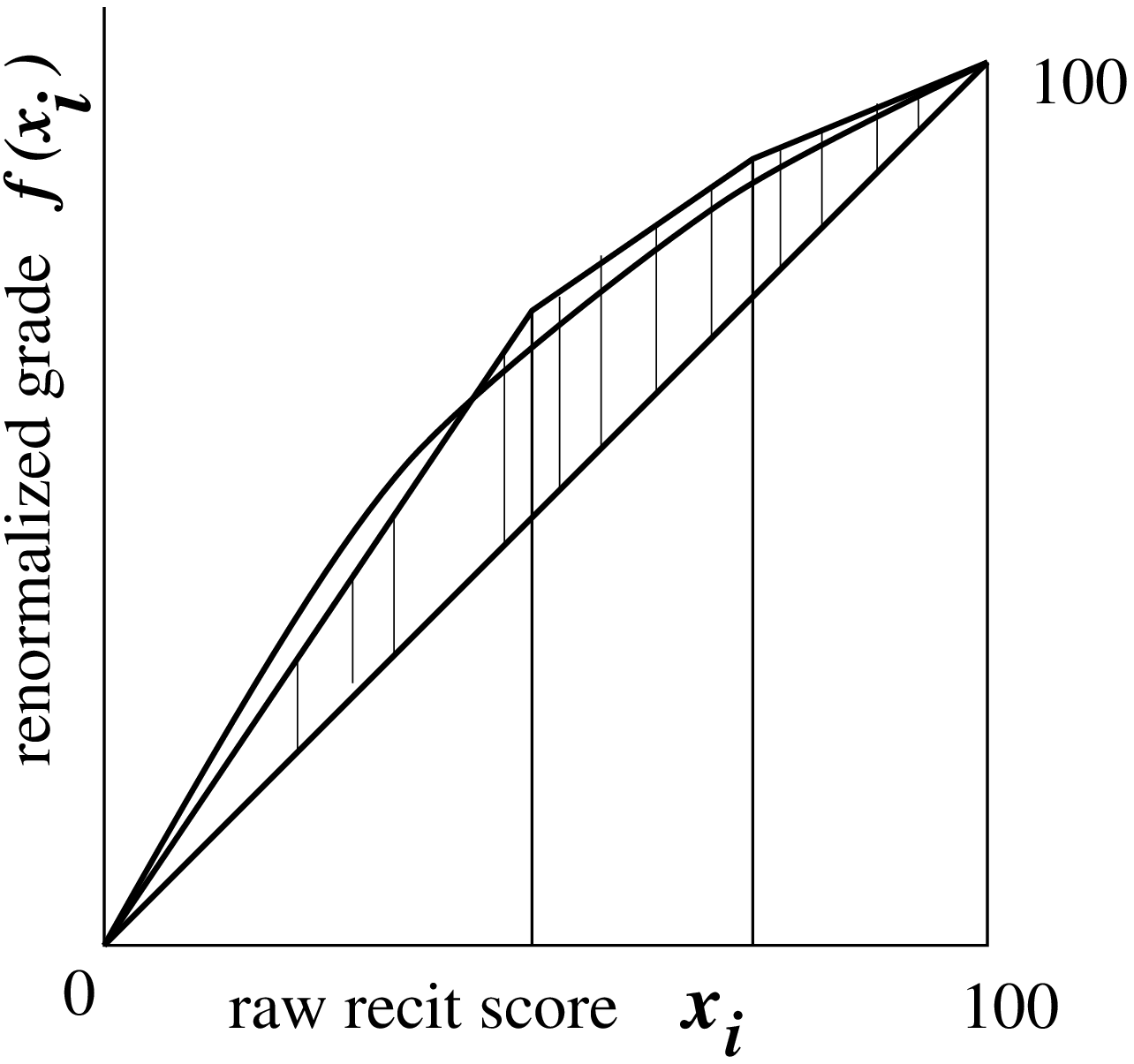}\\
{Possible Renormalization Functions.} 
\end{minipage}\\[4pt]

The method which has been in use in our Department for many years is this:
The average
score $\langle x_i \rangle$ is mapped to g, and this point is connected
by straight lines to $(0,0)$ and $(100,100)$. Clearly, as long as $g$
is sensible, $g\in(0,100)$, this function satisfies (1--3), 
but unfortunately it doesn't satisfy the original criterion:
\begin{description}
\item[0:] $\langle f(x_i) \rangle = g$.
\end{description}

Every semester, in the brief interval between the final exam being
given and the due date for the receipt of grades by the registrar, a 
panic ensues when some new recitation instructor discovers that the
computer does not do what she expects it to do.
There then ensues a discussion 
of whether the fact that this method undercorrects the section grades 
compensates the fact that the goal is overly dependent on the section
exam average. The answer is ``not in general'', and in fact the two
faults can work in the same direction rather than cancel.

One of our faculty came up with an mathematically elegant
renormalization function which satisfies requirements (1--3) and sets
and achieves a goal for the average of the logarithm of the grades,
which at first blush seems as sensible a criterion as a goal for the 
average grade. Considering the grades on the domain $[0,1]$ instead of
$[0,100]$, he noted that scaling the logarithms automatically kept the
endpoints fixed, and therefore
$$ f(x) = e^{b\ln(x)}, \hbox{ where } b =
\ln(g)/\langle\ln(x_i)\rangle$$
provides a renormalization function, easily calculable and simple to 
implement. 

While this solution is mathematically elegant, I think it would be a
very poor choice to use.
The trouble with this method is that it gives the greatest weight in
adjusting the class average to the worst students --- in its pure form
this method would give almost everyone in a class a perfect score if
one student got a zero. We could avoid this by putting in cutoffs on who got
included in the average, but still the weakest students included are
determining the renormalization of everyone. And the elegance is lost 
if we apply it only to students in some subdomain.
I don't believe this is what we want.

\bigskip
\noindent{\large \bf So what can we do?}
\medskip

This paper makes suggestions for how each of these problems can be
solved. Neither is very simple, but each can be readily implemented
by computer. I believe my method for determining the goal
for each section has a ``rightness'' about it, although I can't give a
set of assumptions under which it is ``correct''. The method of
determining the renormalization function is somewhat arbitrary,
but it should (well, under reasonable conditions, at least) 
satisfy (0--3), and also weigh students fairly evenly, emphasizing
those in the middle.

\medskip
\noindent {\bf Determining the goal}

\noindent\begin{minipage}[t]{2.75in}
\parindent=1.7em

If each student deserved a recitation grade which was a monotone
function of her exam grade, and if we knew the appropriate
distribution of recitation grades for the full class, the recitation
grade would be simply given by
\begin{equation}
 x_i = r(E^{-1}(e_i)),
  \label{distmap}
\end{equation}
where $r(n)$ is the $n$'th recitation grade in ascending order, and 
$E(n)$ the $n$'th exam grade in ascending order. 

Of course, recitation grades, even from the same instructor, are far
from well defined functions of the exam grades. Other factors affect 
recitation performance differently than exam performance, and are a 
legitimate component of a student's final grade.
Nonetheless, (\ref{distmap}) could be used to estimate what the expected
recitation average would be, averaging over other 
\end{minipage}
\mbox{\ }
\begin{minipage}[t]{2.4in}
\leavevmode{\tiny\phantom{.}}\\
\epsfxsize=2.4in\epsfbox{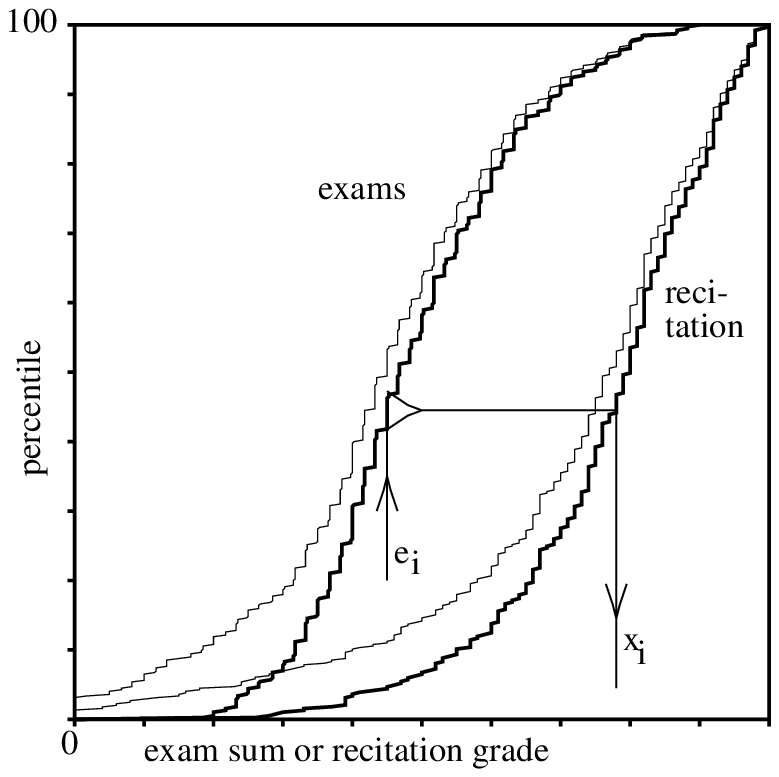}\\
Cumulative histograms of exams and raw recitation grades, for the full
class (thin) and for ``qualified'' students (thick). The goal for each
section is set by averaging the ``expected'' grades $x_i$ for the
qualified students in that section.
\end{minipage}\\[3pt]
factors, were it not for the differences in instructors. 
The function $r(n)$, the ``correct''
distribution of recitation grades, is undetermined, but I think its exact form
is not terribly important --- the crucial issue here is fairness, 
and as long as the distribution has a reasonable spread, it doesn't 
much matter where its mean is, because that will just affect where 
grade dividing lines are set\footnote{Some may be disturbed by my
  refusal to discuss cosmetic issues, violating such rules as ``it's
  okay to raise students' grades but not to lower them''. Such
  instructors could use my method, find the maximum amount any
  student's grade was lowered, and simply add that grade to every
  student. Presumably the grade cutoffs will also be raised by that
  amount, so not a single grade will be affected, but everyone might
  feel better.
}. I think we should use the actual raw
distribution of the full class as our function $r(n)$.

The above discussion applies best if we restrict our attention to
students who have taken all exams and who have attended some
reasonable minimal number of recitation sections. Only those students
are used to determine the functions $r(n)$ and $E(n)$, and 
the section average goals, and thus the renormalization functions for
each section.

Thus I suggest taking as the goal for section $S$ the average
$$ g =  {1\over n_S}\sum_{i\in S} r\left(E^{-1}(e_i)\right),$$
where $n_S$ is the number of qualified students in section S, with the
sum only over the qualified students. The minimum number of recitation
grades required for qualification can be specified by the course leader.
Of course all students will have their grades calculated using the 
renormalization function for their section, whether or not they were
included in determining the function.

\medskip

\noindent {\bf The Renormalization Function}

  The renormalization function needs to be a monotone function such as
  the possibilities shown above. 
  It ought to keep close grades close and large differences
  large, so the slope ought to stay as close to 1 as possible, and
  nonetheless distribute the required modification to the sum of the
  grades reasonably uniformly. I see no justification for 
  requiring continuity of the derivative  $f'$, and therefore no
  objection to using a piecewise linear function.

  It would be straightforward to use two 
straight line segments, perhaps joining at the average raw grade for
the section.  The  
student at the break point would be moved further than $g$, because
other students will be moved less, and thus the average will be moved
by less than the student who had the average raw grade. 
If we use three line segments, which I recommend, this will still be
true but the maximum movement required is less.

If $f(x)$ is given by three continuous line segments, the parameters
are the positions, $x$ and $f(x)$, of each of the two breaks. Four
parameters are a lot to fit, and to make things easier I fix the
break points in the domain $d_1$ and $d_2$ to subdivide the section
into three approximately equal numbers of students. This leaves two
parameters, $f(d_1)$ and $f(d_2)$, constrained by the 
requirement that the average come out right. This is one linear
condition on the two parameters, with coefficients given in terms of
the goal $g$, the numbers $n_L$, $n_M$, $n_R$ in the left, middle and
right thirds of the class, and the sums $s_L = \sum_{i\in L} x_i$,
$s_M$ and $s_R$ of the raw grades in each of these subsets. These are
all easy to accumulate.

We still require one further condition to specify the parameters,
which is set by minimizing some measure of how badly the
renormalization is distorting the recitation grades. I have considered
and implemented two
different measures. The simpler, (``minimax'') minimizes the
maximum change, in absolute value, in a student's grade. This requires
the slope of the middle line segment to be 1, and is straightforward
to implement. I also considered 
minimizing the deviation of the slopes from 1, because two students who
got nearly the same raw grade should not get very different
renormalized grades, or vice-versa. I measured the ``badness'' by the
sum, over the three segments, of the square of the sum of the slope
and its reciprocal. This treats $f$ and $f^{-1}$ on the same footing, 
declaring either a zero or infinite slope to be infinitely bad. Both 
methods work well and fairly easily, though the second requires a
numerical root finder to find the minimum. I call this the ``slope'' method.
Each method will lower the 
grades of some students, the difference perhaps best illustrated by
the effect in one section with inflated grades in my course.
The minimax method kept the maximum change to 7 points, but
lowered a 92 to an 85, nearly doubling the points lost. 
The slope method
lowered one student by more, 10 points, but his grade was a 76 lowered to a
66, perhaps less dramatic, 
while the student with a 92 was lowered only to an 87. 
An even more extreme example is provided by a test run halfway through
the current term (see below), in which the minimax method lowered
grades by 16\%, including dropping a 98\% to a 82\%, while the slope
method dropped the 98\% to 96\% but dropped a 90\% to 68\%. 
I find the slope method results less troublesome, but others might disagree.

\bigskip
\noindent{\bf Results}

Both methods were used in a third term Engineering class with 440
students and 15 sections taught by five recitation instructors, in the
fall of 1996, with the final grades utilizing the minimax method.
In the spring of 1997, in the fourth term of that sequence, with 405
students in 13 recitation sections,
the slope method was used to determine the
final grades. None of the instructors in either semester
complained of the results, which looked as reasonable as could be 
expected given the widely divergent grading of my recitation
instructors. Examples from last fall are shown in Figure 4.

\noindent\begin{minipage}[t]{2.75in}
\parindent=1.7em

Both methods were used at the halfway point of the semester this spring, 
primarily to point out to one instructor what will happen to
his students. He was giving highly inflated grades; one third of the
class had 98\% and above, and two-thirds had 90\% and above, while the
goal for this section was 80.4\%. My intention was to have the 
instructor change his ways so that the final scores would need only
moderate renormalization\footnotemark. 
But this exercise  also provided a strenuous test of
the method. 

Although in principle either\break
method can fail if asked to change\break
an average excessively, at no point did that happen, 
even in the extreme case of the section at midterm described above.
For safety, the program graphs the renormalization function for each
section so that sick solutions will be spotted. By sick solutions, I include
non-monotonic ones. Such a solution can in principle arise in the minimax
\end{minipage}
\mbox{\ }
\begin{minipage}[t]{2.5in}
\leavevmode{\tiny\phantom{.}}\\
\epsfxsize=2.5in\epsfbox{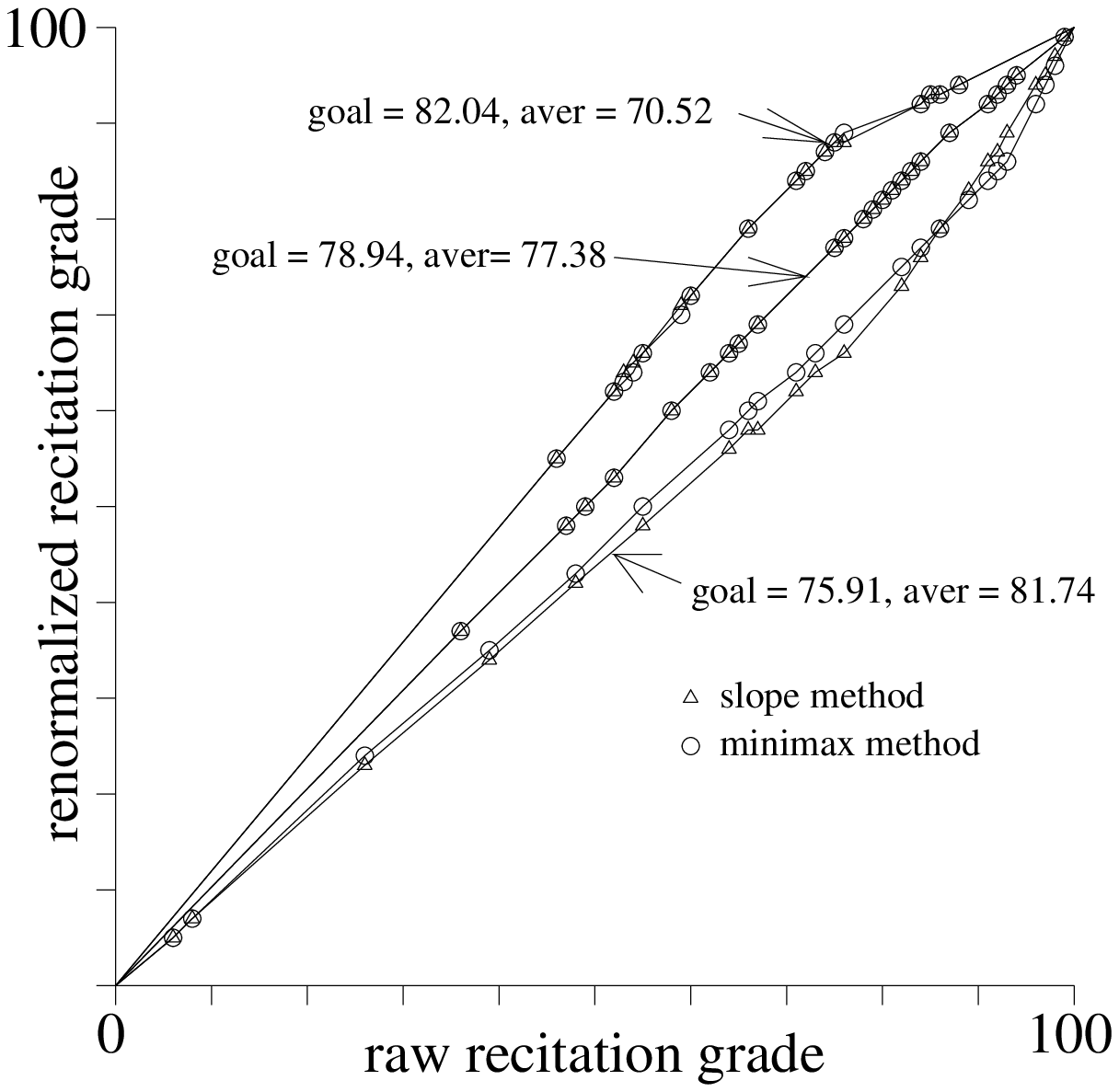}\\[8pt]
{Figure 4. Examples of section Renormalizations.}\\[4pt]
\hskip 1em
Three sections from last fall's class, renormalized by each of the 
two proposed schemes. The sections are the most inflated, the most
harshly graded, and the one requiring the least renormalization. Only
for the most inflated section are the results of the two methods 
significantly different.
\end{minipage}\\[4pt]
algebra,\footnotetext{It was partially successful in doing so.}
and also if the minimization 
routine used by the slope method jumps over the infinite badness point when
a slope goes to zero. I believe such problems are not likely to arise 
in practice, and this did not happen in any of sections in my course.

\penalty -9000
\noindent{\bf Summary}

Thus we now have available methods for ensuring that the recitation
grades assigned to students can be assigned fairly, taking account of 
different average abilities among the sections while eliminating the
effects of different grading standards among the instructors. 
Details of the method can be obtained from the
author\footnote{shapiro@physics.rutgers.edu,~or~\nobreak
http://www.physics.rutgers.edu/$\sim$shapiro/grades.html .}.

\bigskip
\noindent{\bf Acknowledgements}

I wish to thank Frank Zimmermann for several interesting 
discussions, which reemphasized for me the need to think out a
solution to this problem. These conversations also made clear that the
problem was nontrivial, not only to solve but, even more so, to
state. I also thank 
Juliet Shaffer for reading this document and making several helpful
suggestions for clarifying the presentation.

\end{document}